# Dynamics of Acoustically Levitated Ice Impacts on Smooth and Textured Surfaces: Effects of Surface Roughness, Elasticity, and Structure

*Adam McElligott, André Guerra, Alexandre Brailovski, Shashini Rathnayaka, Xiaodan Zhu, Alexia Denoncourt, Alejandro D. Rey, Anne-Marie Kietzig, Phillip Servio\**

Department of Chemical Engineering, McGill University, Montreal, Quebec H3A 0C5, Canada

*<u>phillip.servio@mcgill.ca</u>

ABSTRACT

The effect of atmospheric ice collisions on exposed outdoor infrastructure is a rarely explored surface design challenge in the lab setting. Through acoustically levitated ice formation and subsequent release onto a controlled area, this study introduces a third class of ice-countering system beyond de- and anti-icing: ice-impacting. By subjecting stainless steel 316 (SS), epoxy resin-coated (ER), and laser-textured (LT) surfaces with known surface roughness, hardness, and structural characteristics to 40 ice droplet impacts each, the effect on surface properties and their effect on solid-solid interfacial impact dynamics, in turn, was examined using a novel analysis framework based on fundamental conservation laws. For the velocities experienced in this study, the impacts did not affect the surface properties; they were consistent after each impact. Elasticity was the most significant factor in droplet behavior: the ER surface exhibited rebounding for 78% of impacts (important for moving surfaces). Surface roughness also played a role, particularly for droplets with rotational motion, as immobilization occurred for 66% of impacts on the rougher LT surface. However, the nanostructures on that textured surface resulted in droplet redirection perpendicular to the surface directionality (critical for stationary surfaces). In contrast, the other surfaces saw no change or no consistent change in rebound angle. Elasticity also affected momentum retention, where the ER surface had a translational restitution coefficient of 0.32 compared to 0.17 for the two stainless steel surfaces. Surface roughness was the predominant aspect of energy retention: the LT surface had a translational-to-rotational energy transfer coefficient of 0.07 (0.23 for the smoother surfaces), resulting in an overall energy retention coefficient of 0.09 compared to 0.28 for the SS and ER surfaces on average. The results show that when designing ice-countering surfaces, the interplay between elasticity, roughness, and structure at a dynamic solid-solid interface must be considered to optimize their effectiveness.

## 1. INTRODUCTION

Ice development on the wings and fuselage of aircraft has been a tenacious engineering challenge concerning safety and efficiency for decades.[1-3] Glacial accretion on these primary structures increases aircraft weight, which adds aerodynamic drag, leads to more fuel consumption, and can result in equipment failure.[4, 5] Beyond the aviation industry, the icing of outdoor infrastructure exposed to wintry conditions is unavoidable. Ice can damage power transmission lines or telecommunications instruments and reduce the transmissivity of solar panels.[6-8] As a result, there has been significant research interest in designing and testing ice-countering surfaces, particularly over the last decade, to mitigate the negative effects of atmospheric icing and thus lower the environmental impacts of these effects.[9, 10]

Previous strategies to reduce or eliminate ice accumulation through surface design have largely fallen into two classifications. First, there are anti-icing surfaces, which focus on preventing ice formation, usually by reducing the adhesion strength of water.[7] Water repellency is an important characteristic for limiting the presence of glaze ice from supercooled atmospheric water that freezes a short time after impacting an exposed surface.[11] Testing anti-icing properties usually involves dynamic supercooled droplet impacts onto surfaces that are often biomimetic and passively produce a non-wetting state.[5, 12] Water sprays or icing wind tunnels are commonly used for these tests producing small droplets that frost the surface or much larger ones mimicking rain.[5, 6] Second, there are passive ice-shedding or de-icing surfaces, which focus on lowering the adhesion strength of ice so it can be removed more easily.[7] This type of ice-countering is considered passive because, even if an energy input is required to dislodge the ice, the change in adhesion strength does not require energy. This is compared to active forms of novel ice removal requiring additional energy inputs, such as integrating heating pipes or using a plasma heat knife.[10,

[13-15] De-icing additionally focuses on removing rime ice from atmospheric water freezing immediately upon impact.[11] To that end, testing these pagophobic surfaces involves static freezing, where water is placed on a surface, frozen in place, and then manipulated toward removal.[2, 5, 6, 8, 10-12, 15] These tests examine how different surface properties can be modified to (1) lower surface energy (e.g., using smoother surfaces that have a lower thermodynamic work of adhesion), (2) limit water-solid contact area (i.e., reduce the ice-substrate interlocking that occurs when water penetrates the surface roughness then freezes), or (3) change the mechanics of ice structural deformation.[2, 7] Notably, there has also been a recent trend in ice-shedding surface studies of using biomimetic surfaces with laser-induced micro- or nano-structuring that are hierarchical and have a defined orientation.[2, 5, 7]

However, the pre-existing solid ice particles that form in clouds can also be a source of ice deposition.[10] All these previous studies have examined systems where water is initially liquid and, therefore, have not examined what can be considered a third ice-countering class: surfaces that reduce ice-impacting effects (i.e., dynamic solid-solid interfaces). These effects are important to investigate as solid ice impacts have a greater probability of modifying the surface compared to liquid impacts or static freezing under multiple icings. In other words, while surface characteristics may modify the dynamics of ice impacts, these impacts may change the properties of the surface in turn. Ice-impacting has yet to be investigated due to the limits of creating atmospheric-like ice (i.e., ice formed without a solid surface present) at the laboratory scale in a system that can also then release that ice onto a surface with a highly reproducible droplet volume, shape, velocity, and angle of incidence (AOI). Recent advancements in the crystallization of acoustically levitated fluids now allow atmospheric-like ice to be made in the lab in an open system, permitting that ice to fall directly onto a chosen surface without interference.[16, 17] Using novel levitated cooling systems avoids the issue of physical manipulation, allowing for greater droplet consistency, and the open configuration of newer levitators allows for direct observation of effects within and around the acoustic field.

This study aims to examine how surface roughness, elasticity, and structure affect the impact dynamics of atmospheric-like ice on smooth and textured surfaces. Additionally, it will investigate how multiple ice impacts over a small area consequently modify these properties. To this end, the use of a common construction material (i.e., stainless steel 316), an epoxy resin found in the aviation industry, and a regularly nano-patterned surface will be explored. These substrates were chosen as they cover a range of characteristics, and understanding their ice-countering applications can be critical to the industries in which they are employed. Moreover, they are commonly investigated for anti-icing and de-icing properties by other researchers and thus complement existing work in the field. No analytical framework currently exists for describing this new class of ice-countering surface effects. Therefore, this study will endeavor to create a framework primarily using fundamental laws of nature, namely energy and momentum conservation, that can be broadly applied in future research.

This study is the first to create and employ a system that can reproducibly form ice without a solid surface and cause it to impact a controlled area. Indeed, examining levitated ice impacts and post-solidification applications of levitated crystallization systems is entirely novel. Finally, the analytical approach to quantifying and qualifying such a system and impact effects on different surfaces is, by nature, a new methodology by which ice-countering effects may be approached in further investigations in the field.

## 2. MATERIALS AND METHODS

2.1 Surface Characterization

Three surfaces were chosen to reflect materials commonly utilized in ice-countering applications that also provided an opportunity to examine some of the most relevant characteristics for reducing ice-impacting effects. The control surface (SS) was made from stainless steel 316 with a relatively low initial roughness (0.046 ± 0.004 µm) as it is a common construction material used in both aviation and infrastructure. The second surface consisted of the control surface coated with an epoxy resin (ER) as epoxy composites are often employed in aeronautics for corrosion and impact resistance and reducing overall aircraft weight.[4] The ER surface was specifically selected to initially have similar roughness and wettability to the base surface, though with a lower hardness and greater elasticity. The final surface involved laser treatment (LT) of the control surface. More precisely, the LT surface was decorated with laser-induced periodic surface structures (LIPSS) of highly homogenous nanoscale ripples. This surface was chosen for its similar hardness (within 5 MPa Vickers hardness) to the control surface, though with greater roughness, lower wettability, and textured compared to smooth topology. LIPSS-decorated surfaces were used in a previous ice-shedding study by Wood et al. (2022).[7] As such, examining these surfaces was also important so that a type of surface topology which has been part of anti-icing studies could equally be investigated for ice impact effect reduction. Note that as the rippled structure of the LT surface is directional, some preliminary testing was required to determine its optimal orientation for data acquisition, though this orientation did not affect the results. Each surface had a 1 cm x 1 cm contact area for ice impacts that was monitored throughout the impact experiment described in the next section.

To measure the areal roughness of the sample surfaces, laser scanning profilometry was completed with an Olympus LEXT OLS5000-SAF laser microscope (LMPLFLN 50x LEXT lens). Three replicates were completed before the impact experiment and five after. For the SS and LT surfaces, the same sample was used in the before and after analysis. However, as resin layers are transparent by nature, the ER samples had to be platinum coated prior to analysis. As such, separate ER samples were used to obtain the pre-impact data such that the external platinum coating would not affect the ice impact results. In other words, to avoid further modifying the surface when determining roughness, the ER samples used for post-impact surface measurements were not the same as those used for the pre-impact ones. A micro-Vickers hardness tester with a 4.9 N load was used to measure the hardness of sample surfaces. Since the tester physically indents the surface, it was necessary to use separate samples for the pre- and post-impact data for all three surface types. This method failed to give a reliable reading for the ER samples, which can be attributed to their tendency towards brittle fracture. The technical data sheets for the cured epoxy resin report hardness values in the 144 to 152 MPa range. Further discussion can be found in the results section.

Dynamic contact angle measurements were carried out using a custom-built contact angle goniometer and dynamic contact angles were determined with the ARCA finder software.[18] As this data collection is non-destructive, the same samples were used before and after impact for analysis, regardless of surface type. Only ER samples showed distinctive receding contact angles. Therefore, for comparison, only the advancing contact angle (ACA) will be reported in this study. Finally, scanning electron microscopy was used to observe the surface topography of the samples before and after impact. The data was obtained using a FEI Quanta 450 FE-ESEM instrument. For the SS and LT surfaces, identical samples were used for pre- and post-impact data. For consistency, the ER surface used for the initial surface roughness and hardness tests was also used to examine the initial topology (prior to platinum coating). A separate ER sample, which would have only undergone the contact angle measurement, was used in the impact experiment and for post-impact measurements.

2.2 Acoustically Levitated Ice Impacts

In addition to the necessary and novel components, the experimental setup described in this section includes the TinyLev acoustic levitation device first developed by Marzo et al. and the cryogenic freezing system ("cryogun") developed by McElligott et al. for levitated crystallization.[16, 17, 19] Further information on the technical aspects of levitation and ice morphology during the freezing process can be found in these sources. A simplified version of the complete setup is presented in **Figure 1**.

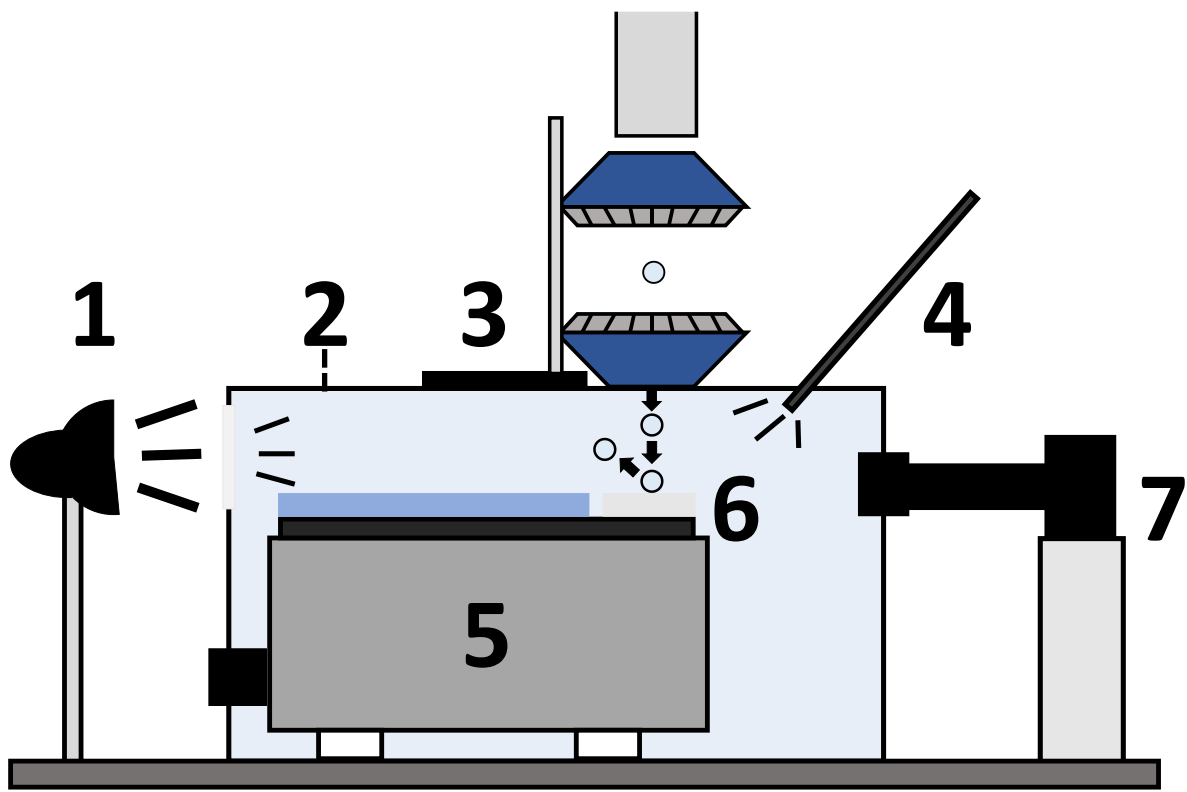

**Figure 1.** A schematic of the experimental setup (simplified). The setup was primarily comprised of a diffused backlight source (1), an insulating container (2), a levitated freezing apparatus (3), a front LED light source (4), an insulated cold plate (5), an impact surface (6), and an ultra-high-speed camera (7). The droplet rebounding direction is arbitrary and not indicative of behavior.

The function of the setup was to create an ice droplet of consistent volume and geometry that could be dropped within a precise area on a cooled surface, the impact upon which could be recorded in detail. To that effect, the droplet was suspended using a TinyLev acoustic levitator (marked as 3 in the figure) operating at 10 V and 40 kHz, which created an acoustic field like those used in previous studies.[16, 17] A bespoke stainless-steel cryogun (also 3) was placed directly above the TinyLev and centered on the axis of levitation. When filled with liquid nitrogen, the cryogun cools to -55 °C and forms a cooling stream capable of initiating and maintaining crystallization within the acoustic field. Placing a droplet in the central node of the field (i.e., equidistant from the acoustic transducer rings above and below) for the freezing process resulted in a consistent 1.61 ± 0.13 mm$^3$ ice droplet volume. This node was chosen as it is considered the most stable and resulted in a drop distance of exactly 20 cm. Therefore, the 1 x 1 cm$^2$ impact area of the surface (6) was also centered with the cryogun and axis of levitation, though the surfaces had a greater area than this so that they could be handled more easily. The surfaces were placed on top of a thermoelectric cooler module (5) from TE Technology consisting of a Peltier cold plate (CP-200TT), a temperature controller, and a power supply. A general-purpose MP-3193 thermistor (operating range of -20 °C to 100 °C) monitored cold plate temperature, and a LabVIEW$^{TM}$ Virtual Instrument was used to set controller parameters. As the cold plate surface was much larger than necessary, most of it was insulated with a 1-inch-thick piece of Styrofoam. This is the larger blue rectangle on top of the cold plate in **Figure 1**. The cold plate was contained by a custom-built cold chamber (2) also made from Styrofoam. This chamber allowed for faster cold plate cooldown and maintained colder temperatures for the entire downward trajectory of the droplet. The chamber had five openings: one for imaging, two for lighting, one for the droplet (the TinyLev and cryogun were placed on top of the chamber), and one for hot air exhaust from the cold plate. The ice impacts were captured via an HSI FASTCAM Mini AX50 ultra-high-speed camera (7) operating at 10,000 FPS and placed perpendicular to the surface. Video playback and analysis were performed at 30 FPS. Finally, additional lighting was necessary for optimal visualization. A Schott KL 2500 fiber optic LED (4) was used to light the part of the droplet facing the camera, while a lamp (1) perpendicular to the camera was the primary lighting source and was used to delineate the droplet edges. Due to excessive back illumination, two thin, white fibrous screens were placed in front of the latter source to diffuse light.

Prior to an experimental run, it was necessary to cool the system, which was done mainly through the cold plate. After placing the impact surface on the cold plate, all cold chamber openings were closed via their respective utility, and the plate was set to -15 °C. This temperature was chosen to replicate the conditions in which atmospheric ice could form and has been used in previous de-icing studies.[2] In addition to surface cooling, the cryogun was also cooled prior to starting the experiment by filling it with liquid nitrogen and allowing two minutes for it to dissipate.[16, 17] After two minutes had elapsed the system was sufficiently cooled, and a run would begin by turning on the levitator, placing a droplet of liquid water in the specified node, and filling the cryogun with liquid nitrogen. After one minute had elapsed and the cooling stream was no longer strong enough to dislodge the droplet, the cryogun was placed over the TinyLev, and the droplet was given one minute to freeze. Previous studies have shown that this length of time is more than enough to result in a completely frozen droplet.[16, 17] After the one minute had elapsed, three events occurred simultaneously: the cryogun was removed, the high-speed camera began recording, and the levitator was turned off. These allowed the droplet to fall accurately towards the impact area and for the impact to be captured. The camera was set to record for no longer than two seconds, which was adequate for a 20 cm drop. After a run, the droplet would occasionally remain on the surface. As such, it was necessary to carefully remove the impact surface, allow any ice on that surface to melt, and dry the surface with a stream of nitrogen gas. Once it was completely dry, the impact surface was placed back in the same position on the cold plate, allowed to cool down again, and the next run could begin. After all the runs were completed, the Photron FASTCAM Viewer 4 software was used to determine droplet dimensions, as well as the translational and rotational velocities during impact and rebounding. The impact and rebounding angles were also measured using this program. For accuracy, translational (impact) velocities and all angles were consistently measured starting from when the droplet was 3 mm from the surface. Each of the three surface types was subjected to 40 impact runs. However, due to rare variations in the precise impact location, the droplets could impact beyond the focal range of the camera, which was less than 1 cm, and those videos were not sufficiently clear to acquire data. In other words, there were infrequent runs where velocities and angles were not possible to discern as the impact was too close to or too far from the camera, even if it was within the 1 x 1 cm$^2$ impact area. Therefore, while post-impact surface characteristics are based on 40 impacts, the remaining analysis was performed on 36, 32, and 38 runs for the SS, ER, and LT surfaces, respectively.

2.3 Analytical Framework

It was necessary to devise a new analytical framework to analyze ice impact dynamics and how they change depending on surface properties. This framework was primarily derived from known fundamental conservation equations of momentum and energy; mass was assumed to be conserved. Simplifying the system to only operate in two dimensions, which was acceptable to assume based on the results, the critical droplet states to characterize the impact are given in **Figure 2**. These were pre-impact (A), which was defined by an initial lateral velocity $V_1$ and an initial rotational velocity $r\omega_1$, impact (B), where energy and momentum exchange could occur between the droplet and surface, and post-impact (C), defined by a rebounding velocity $V_2$ and a rebounding rotational velocity $r\omega_2$. The droplets could have initial rotational velocities due to turbulence in the acoustic field, which may be similar to atmospheric turbulence due to wind or turbulence in the boundary layer at the surface of an aircraft wing. Note that the figure suggests that the impact and rebound angles were equivalent. However, this was not necessarily the case, and the impact or rebound angles were also considered when determining $V_1$ and $V_2$.

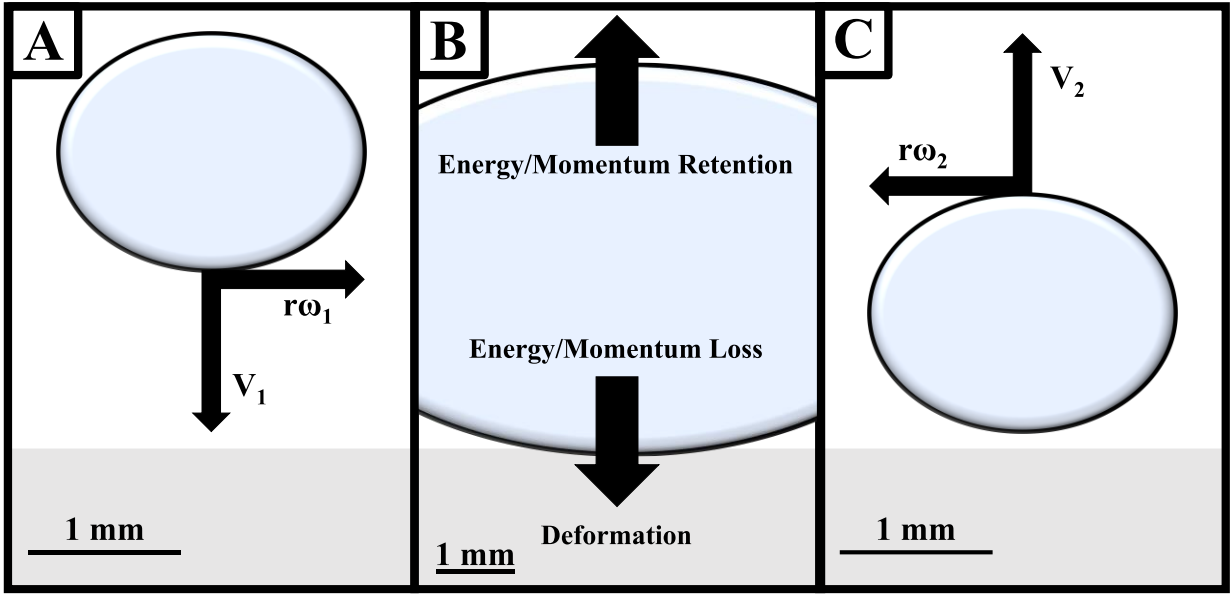

**Figure 2.** The three critical states for impact analysis were pre-impact (A), impact (B), and post-impact (C).

With four measured velocities, the system could be defined by a system of four equations, which were then simplified into four dimensionless coefficients for analysis. These coefficients were (1) the translational restitution coefficient, $\varepsilon_T$, which measures translational momentum retention, (2) the rotational restitution coefficient, $\varepsilon_R$, which measures rotational momentum retention, (3) the retained energy coefficient, $C_{Ret}$, which measures the energy retained by the droplet, and (4) the transferred energy coefficient, $C_{Tr}$, which measures the energy transferred from translational to rotational upon impact. The equations defining the four coefficients are found below. The derivations for these equations are present in the Supplementary Information.

$$\varepsilon_T = \frac{V_2}{V_1} \tag{1}$$

$$\varepsilon_R = \frac{\omega_2}{\omega_1} \tag{2}$$

$$C_{Ret} = \frac{V_2^2 + (r\omega_2)^2}{V_1^2 + (r\omega_1)^2} \tag{3}$$

$$C_{Tr} = \left|\frac{(r\omega_2)^2 - (r\omega_1)^2}{V_2^2 - V_1^2}\right| \tag{4}$$

However, while this system accounts for the energy lost upon impact, it does not differentiate between thermal losses from friction and losses from deformation of the surface. Therefore, a fifth equation was required to determine the deformation energy coefficient, $C_{Def}$, which measures the percentage of the droplet's lost energy attributed to surface deformation. Moreover, it would also be necessary to incorporate a fifth measurable variable to have a consistent solution to the system. Starting from the Young-Dupré equation for the work of adhesion, the equation below was determined which uses the measured contact angles before and after the 40 impacts, $\theta_1$ and $\theta_2$, as well as the impact area, A, and the liquid surface tension $\sigma_l$. A derivation can once again be found in the Supplementary Information.

$$C_{Def} = \left|\frac{A\sigma_l(\cos\theta_2 - \cos\theta_1)}{\frac{1}{2}m[(V_2^2 + (r\omega_2)^2) - (V_1^2 + (r\omega_1)^2)]}\right| \tag{5}$$

This system of five equations can define the ice impact dynamics entirely and is used in the next section to determine how different surface characteristics affect momentum and energy transfer and establish where losses may arise.

3. RESULTS

The results are divided into two sections. The first details the effects of repeated ice droplet impacts on the material characteristics of the surfaces. The second examines the dynamics of the ice impacts (i.e., how surface properties affect the ice-substrate collision in turn). A discussion of the influence of ice particle shape is available in the Supplementary Information.

3.1 Effects of Ice Impacts on Surface Characteristics

First, the effect of 40 ice droplet impacts on the surface roughness was examined. The results are presented in **Table 1**. These results show that the SS and ER surfaces have similar surface roughness and are fairly smooth, while the LR surface is a full order of magnitude rougher. However, there was no significant post-impact change in surface roughness in all cases. This could be because 40 impacts are insufficient to deform the surface and change its roughness properties. Similarly, the drop distance was 20 cm, resulting in an average impact velocity of 1.36 ± 0.03 m s$^{-1}$. This velocity may not have been great enough to affect the surface, noting that ice impacts on aircraft and outdoor surfaces would likely have more considerable velocities than what was examined in this study. It is recommended that future ice impact studies attempt more impacts at higher velocities for an additional degree of realism.

**Table 1.** Pre- and post-impact surface roughness, Vickers hardness, and advancing contact angle (ACA) of each surface type with their 95% confidence intervals in brackets.

| | Surface Roughness (μm) | | Hardness (MPa) | | ACA (°) | |
|---|---|---|---|---|---|---|
| Sample | Pre | Post | Pre | Post | Pre | Post |
| SS | 0.046 (0.004) | 0.043 (0.003) | 182 (3.6) | 178.2 (5.3) | 85 (2) | 85 (4) |
| ER | 0.027 (0.001) | 0.024 (0.002) | - | - | 78 (0) | 81 (4) |
| LT | 0.376 (0.004) | 0.388 (0.003) | 185 (2.6) | 182.8 (2.9) | 21 (1) | 58 (10) |

Next, the surface hardness for both the SS and LT surfaces was measured and is found in **Table 1**. Again, the hardness of the epoxy resin could not be measured but is likely in the 144 to 152 MPa range. From the table, the SS and LT surfaces have similar hardness. This is logical as both are made from stainless steel. They both have greater hardness than what would be predicted for the more elastic ER surface. Furthermore, there was no change in hardness after 40 ice droplet impacts for the SS and LT surfaces, though it was not anticipated that this would have an effect.

This makes it likely that the ER surface hardness was also not affected, but this cannot be ascertained as it was not measured.

Contact angle measurements are important in this study for understanding the wettability of the different surfaces and are also necessary for the system of equations developed to characterize impact dynamics. The results of the measurements are reported in **Table 1** and show how, initially, the smooth (SS and ER) surfaces have similar advancing contact angles, while the LT surface has a lower one. This indicates that the structure of the LT surface makes it more hydrophilic than the other surface types. However, hydrophilicity is unlikely to influence impact dynamics here as they consist of solid-solid collisions. Instead, a change in contact angle may indicate surface deformation after multiple impacts. Both the SS and ER surfaces have similar pre- and post-impact contact angles, while there is a significant increase in the ACA for the LT surface. This may demonstrate a decrease in wettability due to deformation of the LT surface from droplet impacts. That said, laser-treated surfaces tend to decrease in wettability with time while exposed to ambient conditions, and this decrease is well-characterized in literature.[20, 21] Within the experimental timeframe, it is likely that the decrease in wettability is due to air exposure rather than deformation. To verify this effect, both the impacted and an unimpacted LT surface were measured approximately three months after the initial ACA measurement. The impacted surface had a contact angle of approximately 86º, which is higher than what was presented in **Table 1** despite not having been subjected to further impacts in the interceding time, and the unimpacted surface ACA was 84º, which is nearly identical. Therefore, the change in wettability of laser irradiated surfaces in contact with air over time is the likely cause of the ACA increase rather than surface deformation. Moreover, this suggests that the system that defines impact dynamics can be reduced to four equations for the current study. With no deformation likely occurring, the deformation energy coefficient will always be zero, and it will be assumed that all energy losses can be attributed to thermal losses due to friction.

Finally, we can examine the surface topology. Pre- and post-impact SEM images for the LT surface are presented in **Figure 3**. The images for the other surfaces are available in the Supplementary Information. From the images, it is evident that, unlike the smooth SS and ER surfaces, the LT surface has a directional, textured surface with homogeneous, periodic surface structures. However, the images do not appear to indicate any topological change after the ice impacts. This makes sense as none of the surface properties were affected by the impacts. In short, it can be concluded that the ice impacts did not affect the surface properties investigated here in any significant way. It may be that the velocity was too low or there were too few impacts to have an effect, which would require further examination. In the context of this study, this means that the surface properties were consistent at each of the 40 impacts. The effect of these properties on impact dynamics is explored in the next section.

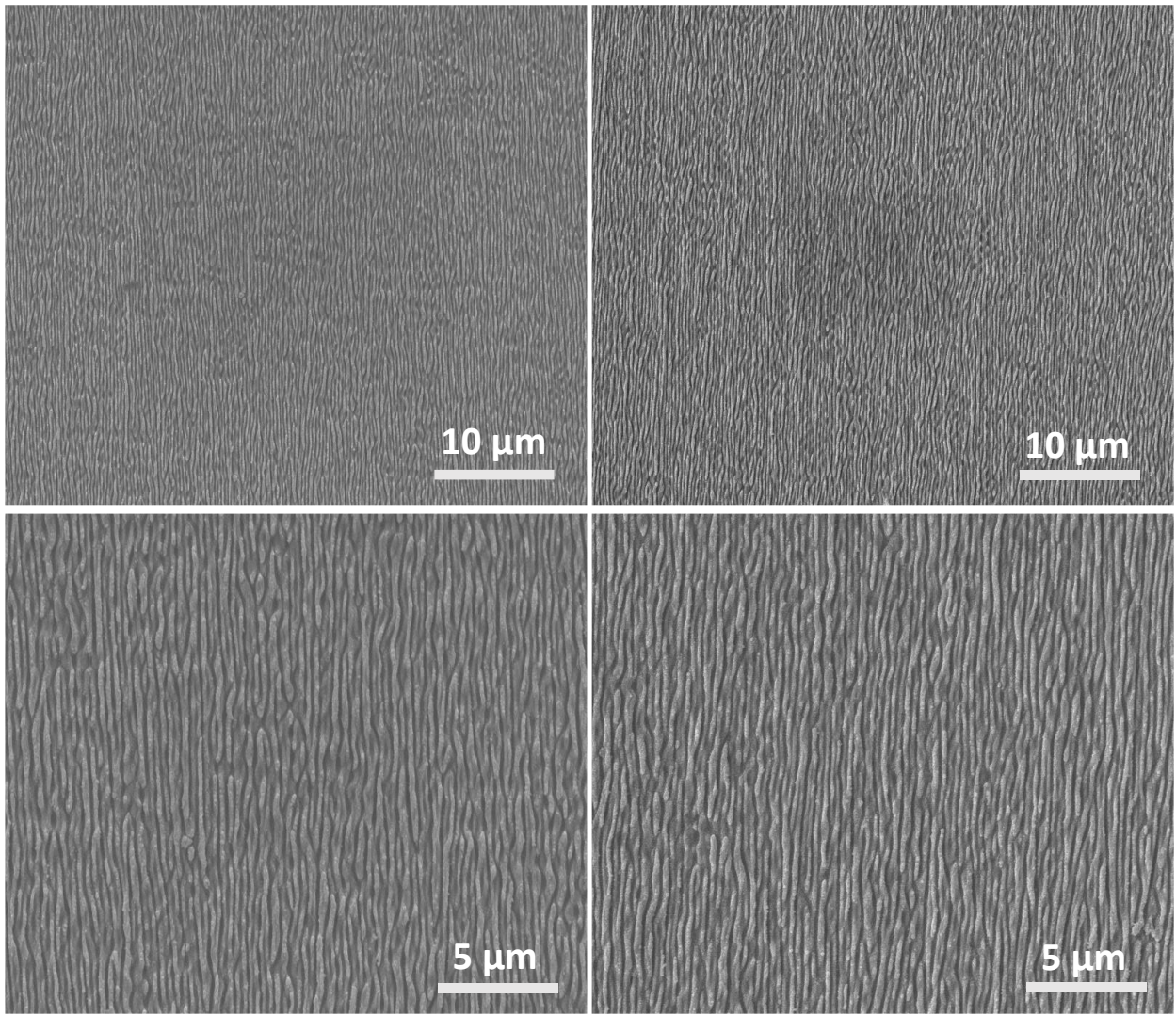

**Figure 3.** Pre- (left) and post-impact (right) surface topology for the LT surface at two magnifications.

3.2 Ice Impact Dynamics

Ice droplets were formed in an acoustic levitator and dropped onto three different surfaces. The dynamics of the ice-surface impacts at the solid-solid interface will be analyzed in this section using the novel system devised earlier in the study based on fundamental conservation equations. Note that this framework assumes that the droplets are highly spherical, which was largely the case as approximately 62% of impacts were from oblate spheroidal droplets. Further information on droplet shape can be found in the Supplementary Information. The initial impact velocity was 1.36 ± 0.03 m s$^{-1}$ with an AOI of 89.63 ± 0.21º. However, it is important to note that because the impact area was small, it was impossible to capture droplets that fell at wide angles; the results are biased toward droplets with approximately 90º impact angles. It may be that droplets impacting at different angles have different effects. This can be explored in future investigations. Again, the average ice droplet volume was 1.61 ± 0.13 mm$^3$, resulting in an average calculated mass of 1.45 mg. Finally, 21 out of the total 106 runs examined had initial rotational motion.

3.2.1 Impact Behavior

Despite similar impact velocities for every run, two distinct impact behaviors were observed and are shown in **Figure 4**. The first behavior was rebounding, while the other was immobilization. Here, immobilization is defined as (1) the particle does not break contact with the surface after impact (i.e., consistent solid-solid interface) and (2) the particle does not complete a full rotation after impact while in contact with the surface. In other words, when immobilization occurred, the post-impact momentum and energy values were zero for the translational and rotational terms. It is possible at higher impact velocities for a third behavior to occur, ice droplet shattering. However, this did not occur for any run in this study.

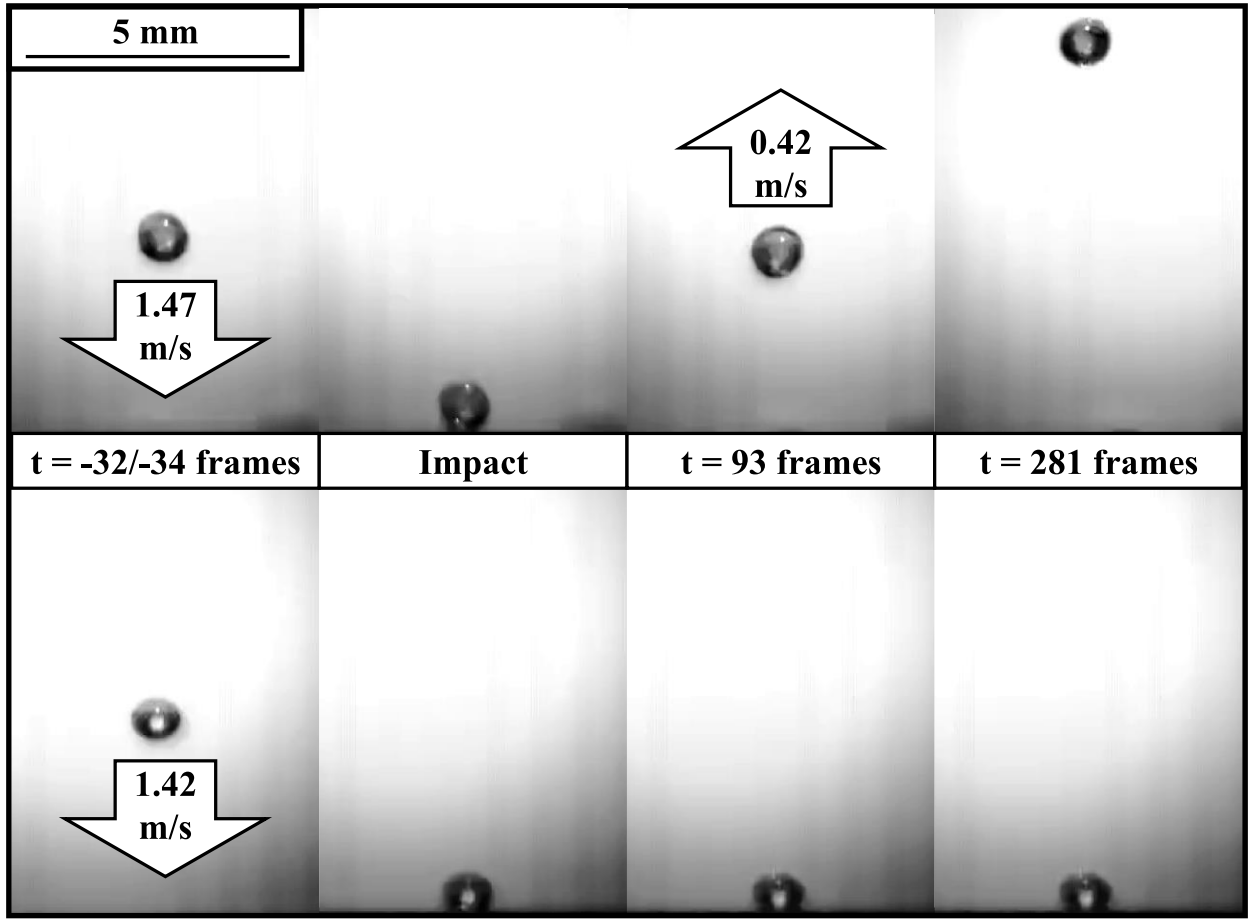

**Figure 4.** Despite similar initial velocities, the two impact behaviors on an ER surface: a rebound (top) and an immobilization (bottom). The left-most and middle-right frames show the droplet at the measurement point, 3 mm from the surface. Times are relative to the impact and the contrast of these images was increased by 40% for visualization.

The percentages of droplets that rebounded were 53% for the SS surface, 78% for ER, and 34% for LT. The higher value for the ER surface makes sense as it is a more elastic substrate and could return more momentum to the droplet. However, due to their comparable hardness properties, a similar solid-solid interfacial momentum exchange was expected between the SS and LT surfaces. Instead, there may have been less rebounding on the LT surface due to increased roughness, and thus friction, immobilizing the droplet (particularly those exhibiting initial rotational motion). Note that the somewhat irregular shape of the droplets may contribute to immobilization. Even if they cannot complete a full rotation upon impact, there will likely still be some rotational motion on an irregular contact area. This could result in sufficient friction for the immobilization of droplets that were not classified as spinning before impact. The LT surface has greater wettability and, unlike this study, most previous anti-icing studies initially have liquid water in contact with the surface, which would result in greater surface penetration by ice compared to a solid-solid collision and manifest different interfacial energy interactions.[2, 5, 6, 8, 10-12, 15] However, there was no evidence that liquid water was present at the LT surface during impact. Moreover, anti-icing studies focus on a single direction parallel to the surface (the critical direction for removing ice from surfaces), whereas there is both perpendicular and rotational motion here. Therefore, in dry systems, the surface roughness may be a greater factor in ice droplet rebound behavior than characteristics like wettability that are often investigated for anti-icing properties.

Examining only rebounding droplets, two of the surfaces (ER and LT) exhibited a directionality to their rebounds. The ER surface consistently (78% of the time) exhibited rebounds directly upward with no change in angle, while 85% of rebounds off the LT surface were perpendicular to the surface directionality (i.e., rebounded to the left or right when examining **Figure 3**). This indicates that the surface topography affected how the droplets rebounded, and giving a direction to the surface likely gives a direction to the rebound. However, surface topography did not appear to affect the rebound angle. Examining the runs across all three surfaces where a non-perpendicular rebound occurred, there was no significant deviation in rebound angle for any surface, flat or textured, with an average of $62.18 \pm 9.59°$. In addition, there was no surface effect on rotational directionality. In other words, if a droplet was rotating both before and after impact, the rotational direction was always the same. Therefore, immobilization on the textured surface suggests that water-repellant surfaces may result in greater ice accumulation on a surface when considering solid-solid impacts, but they can also direct rebounding ice away from that surface. In terms of application, there is some middle ground between materials that reject solid ice via uncontrolled rebounding and those that can consistently and predictably direct some ice away. For example, redirection may be optimal for stationary surfaces: a droplet that rebounds directly upwards, which is essentially not redirected, may end up returning to the surface and result in more accumulation. Redirection may be less effective for fast-moving surfaces, such as those on aircraft, particularly if the rebound angle results in ice being oriented directly towards the surface. In the current investigation, the small impact area is insufficient to determine significant post-impact amassing of ice, and a study on a much greater area should be considered. In short, these results indicate that examining the interplay between surface properties and initial and secondary accretion may be necessary when considering ice-countering surfaces.

3.2.2 Momentum Dynamics

The momentum dynamics of an ice droplet impact onto a dry substrate can be characterized by the translational and rotational changes in velocity: the translational ($\varepsilon_T$) and rotational ($\varepsilon_R$) restitution coefficients, respectively. These values are presented in **Table 2**. Note that as we are examining dynamics and not only behavior, the restitution coefficient values only include the rebounding runs that had a rebounding velocity: immobilized runs had a precise value (of 0) but offered limited information on interfacial momentum exchange in the system. Moreover, as collisions are not perfectly elastic, there will always be some overall momentum loss on impact. However, it is important to distinguish between the translational and rotational components, as some runs exhibited an increase in rotational velocity despite an overall momentum decrease.

Therefore, **Table 2** also divides $\varepsilon_R$ into loss and gain components. As these values consider runs solely with an initial and final rotational velocity (only 8, 11, and 2 SS, ER, and LT runs, respectively), the loss and gain components do not have a strong statistical significance, though they may point to important aspects of system dynamics.

**Table 2.** Translational ($\varepsilon_T$) and rotational ($\varepsilon_R$) restitution coefficients for each surface type, along with their 95% confidence intervals in brackets. In addition, $\varepsilon_R$ is divided into loss and gain components, though these values lack statistical significance.

|  | SS | ER | LT |
|---|---|---|---|
| $\varepsilon_T$ | 0.19 (0.05) | 0.32 (0.09) | 0.15 (0.05) |
| $\varepsilon_R$ | 1.05 (0.58) | 1.14 (0.28) | 0.56 (0.42) |
| $\varepsilon_{RLoss}$ | 0.26 | 0.7 | 0.56 |
| $\varepsilon_{RGain}$ | 1.84 | 1.5 | - |

The polymer resin system has a higher $\varepsilon_T$ than the other two surfaces, which were both made from stainless steel, though generally, all surfaces exhibited significant translational momentum losses. This follows with the elastic properties of the ER surface and with the higher incidence of rebounding on that surface generally, as discussed in the previous section. Indeed, the SS and LT surface values are not statistically different, which may confirm the effect of surface roughness on impact behavior. Therefore, surfaces with lower hardness will return more momentum to an impacting ice droplet. This is relevant to primary ice impact dynamics and potential secondary accretion effects in this system. A surface with greater solid-solid interfacial momentum exchange after the primary impact may have a lower chance of having that same droplet re-impact as it travels farther away, exacerbating small deviations from the 90º impact angle upon rebound. However, higher velocity impacts will need to be investigated as data in the previous section showed that surfaces that return more momentum to a 90º impact at the velocity in this study tend towards identical rebound angles. A representative example of momentum exchange on each surface is shown in **Figure 5**. This figure demonstrates how the translational restitution was not necessarily velocity-dependent and was strongly affected by the material properties. In addition, in these examples, the SS and LT surface impacts resulted in rotational motion, where initially there was none. The ER impact maintained rotational motion, though at a lower velocity.

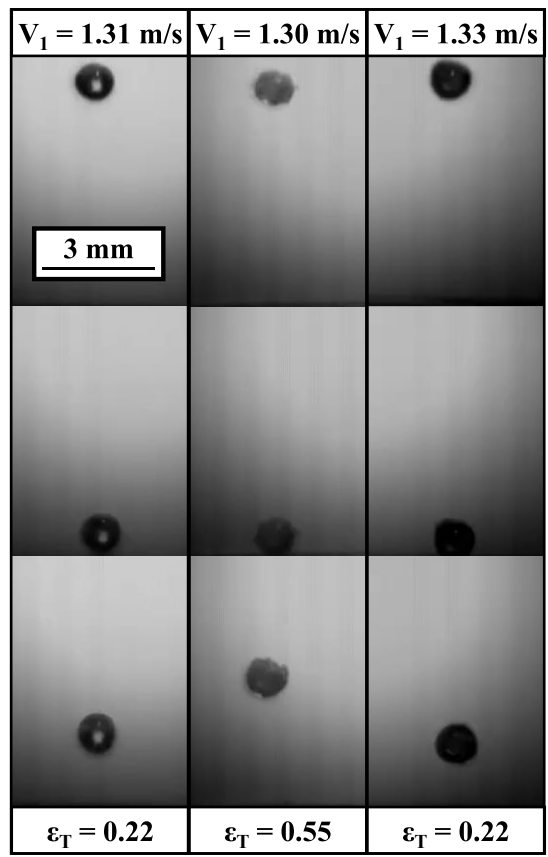

**Figure 5.** Droplet positions 50 frames before and after impact. Translational restitution on either of the SS (left), ER (middle), or LT (right) surface types was not velocity-dependent.

There is low confidence in the results for the rotational restitution coefficient, mainly because there are two distinct behaviors and few runs for the individual behaviors. For example, there was no gain component for the LT runs, which can be linked to surface roughness. Greater surface roughness may eliminate the gain component of the rotational restitution, but this cannot be confirmed by a response that only occurred in less than 20% of runs. The only conclusion that can be drawn is qualitative: the surface with increased surface roughness displayed very little post-impact rotational activity. These results are important as introducing a rotational component adds a frictional force in a different direction than non-rotating impacts, which could result in more wear on surfaces in other systems. However, unless paired with the rotational energy results in the following section, the $\varepsilon_T$ alone cannot be used to make strong claims. Moreover, the current data set does not indicate that rotational velocity is particularly relevant to the previously mentioned applications, even if it is an essential part of defining system physics. Many more runs (i.e., likely 15 times as many as are present in this study) would be necessary to get enough data to increase confidence in these values and determine quantitative differences. Due to the resources necessary to examine such a rare behavior in the system, this kind of specific investigation should be considered for computational studies instead.

3.2.3 Energy Dynamics

The interfacial energy transfer dynamics from an ice droplet impact onto a dry substrate can be characterized by the energy retained on impact and transferred between translational and rotational energy: the retained ($C_{Ret}$) and transferred ($C_{Tr}$) energy coefficients. These values are presented in **Table 3**. Once again, immobilized droplets are not considered here (i.e., these averages consist solely of rebounding runs, which had non-zero values). Including the latter droplets would skew the numbers much lower and obfuscate some of the energy dynamics in the system. Moreover, the transferred energy coefficient is specifically for transfers from translational to rotational energy. In other words, only runs where the rotational energy increased upon impact were considered, though there was always an overall decrease in energy. This is acceptable, as it was never the case that rotational energy decreased upon impact unless the droplet was immobilized. However, if the droplet rebounded twice in the same run, energy was transferred back from rotational to translational. This event was rare, occurring only twice in the 106 runs and only on the ER surface where sufficient energy was retained to demonstrate two instances of energy transfer. More information on the double rebound case is available in the Supplementary Information. Finally, the initial average total energy was 1.45 ± 0.09 µJ. As velocity was measured just 3 mm from the surface, the potential gravitational energy is three orders of magnitude smaller and can be considered negligible. As discussed earlier, elastic potential energy was also negligible, as no surface deformation was observed. Therefore, only kinetic and thermal energy should be accounted for in this system.

**Table 3.** Retained ($C_{Ret}$) and transferred ($C_{Tr}$) energy coefficients for each surface type and their 95% confidence intervals in brackets.

|  | SS | ER | LT |
|---|---|---|---|
| $C_{Ret}$ | 0.23 (0.07) | 0.33 (0.09) | 0.09 (0.04) |
| $C_{Tr}$ | 0.22 (0.06) | 0.24 (0.08) | 0.07 (0.03) |

Examining energy retention, which includes both translational and rotational energy, there were significant energy losses upon impact. There was no statistical difference between the energy retained by the smooth SS and ER surfaces. Though translational momentum retention was greater for the ER surface in the previous section, the behavior is much more similar when factoring in rotational energy. This is further evidenced by similar energy transfer coefficients between the two surfaces. However, the energy retention for the LT surface was significantly lower. Once again, this likely has to do with surface roughness. Droplets impacting the LT surface may have experienced significantly more friction as they spun at the surface and lost more energy. This may also explain why droplets exhibited greater immobilization on the LT surface. Once again, the statistically lower $C_{Tr}$ values for the LT surface indicate that frictional rotational energy losses play a critical role in overall energy losses in the systems. These energy losses are thermal and would have produced a small temperature change. However, the average change was calculated to be $1.8 \times 10^{-3}$ °C, so it would not be visible using an IR camera. Overall, the two coefficients may indicate that smoother surfaces result in greater energy retention upon impact, mainly from an increased ability to transfer energy from translational to rotational at the solid-solid interface. Droplet shape also had an effect here, which is discussed in the Supplementary Information. However, considering the proposed applications, surfaces with greater roughness are not effective in reducing ice-impacting effects. Therefore, the data in this study suggests that smoother surfaces are superior in solid-solid collisions. Solid ice droplets in the atmosphere will likely be spinning due to turbulence. As such, it is necessary that the impacted surface return as much momentum and energy to the droplet as possible to reduce the (1) instances of immobilization and (2) the likelihood that the droplets return to the surface and accumulate. Therefore, when designing ice-countering surfaces, in addition to the considerations for systems with initially liquid droplets, it is critical to examine how, if implemented, the design affects solid droplet impacts. This can help to tune important properties further and fully optimize the effectiveness of the surface.

## 4. CONCLUSIONS

For the first time, atmospheric-like ice was formed in an acoustic levitator, then allowed to fall and impact common substrates with distinct surface roughness, elasticity, wettability, and structure. In addition, a novel five-equation analysis framework was developed to characterize solid-solid impact dynamics in an ice-countering system based on fundamental conservation laws. After 40 impacts in a controlled area, none of the surface characteristics were modified except for the wettability of the laser-textured (LT) surface, which was a result of exposure to air rather than deformation. This meant that the coefficient related to elastic potential energy could be eliminated from the system in this study. It is possible that studies with more impacts or higher velocities would measure some direct influence on surface properties. Upon impact, droplets would either rebound or be immobilized. The smooth surface coated with an epoxy resin (ER) exhibited rebounding 78% of the time, much more than the control stainless steel 316 (SS) or LT surfaces. This likely resulted from a more elastic surface compared to stainless steel, allowing more energy to be returned to the droplet. This was especially the case for droplets with significant rotational motion, which may have been immobilized more by the rougher LT surface. However, rebounds on the ER surface tended to be angled directly upward, which may not be ideal for stationary outdoor surfaces where the ice could return on descent. Rebounds on the LT surface were consistently perpendicular to the directional nanostructured grooves of the substrate, being redirected at an angle of approximately 62°. Therefore, the interplay between elasticity and rebound angle should be considered for dry impacts beyond the anti- or de-icing properties.

Following impact behavior, the ER surface returned 32% of the translation momentum, compared to an average of 17% for the stainless-steel surfaces. Again, hardness was likely the determining factor for this effect. The LT surface's greater roughness may have limited rotational momentum, though there was low confidence in the exact value. Only kinetic and thermal energy were relevant in this study, noting that this included both translational and rotational kinetics. The smoother SS and ER surfaces exhibited greater interfacial energy exchange with an average energy retention coefficient of 0.28 compared to 0.09 for the LT surface. This indicated that surface roughness was a dominating factor in energy retention (confirmed through the translational-to-rotational energy transfer coefficient), which was 0.23 for SS and ER but 0.07 for LT. While the LT surface had greater wettability, these results indicate that surface roughness and elasticity are the critical characteristics when examining initially dry impacts: wettability is not a significant factor at the solid-solid interface. The interplay between these two properties must be considered to optimize the effectiveness of ice-countering surfaces. Specifically, the instances of immobilization and the likelihood that ice droplets return to the surface should be accounted for in addition to other ice-countering features. This study shows that surface properties like elasticity and roughness can be analyzed to determine optimal parameters in future studies in this field.

AUTHOR INFORMATION

**Corresponding Author**

*phillip.servio@mcgill.ca

**Author Contributions**

The manuscript was written through the contributions of all authors. All authors have given approval for the final version of the manuscript.

ACKNOWLEDGEMENTS

The authors would like to acknowledge the financial support from the Natural Sciences and Engineering Research Council of Canada (NSERC) and the Faculty of Engineering of McGill University (MEDA, Vadasz Scholars Program).

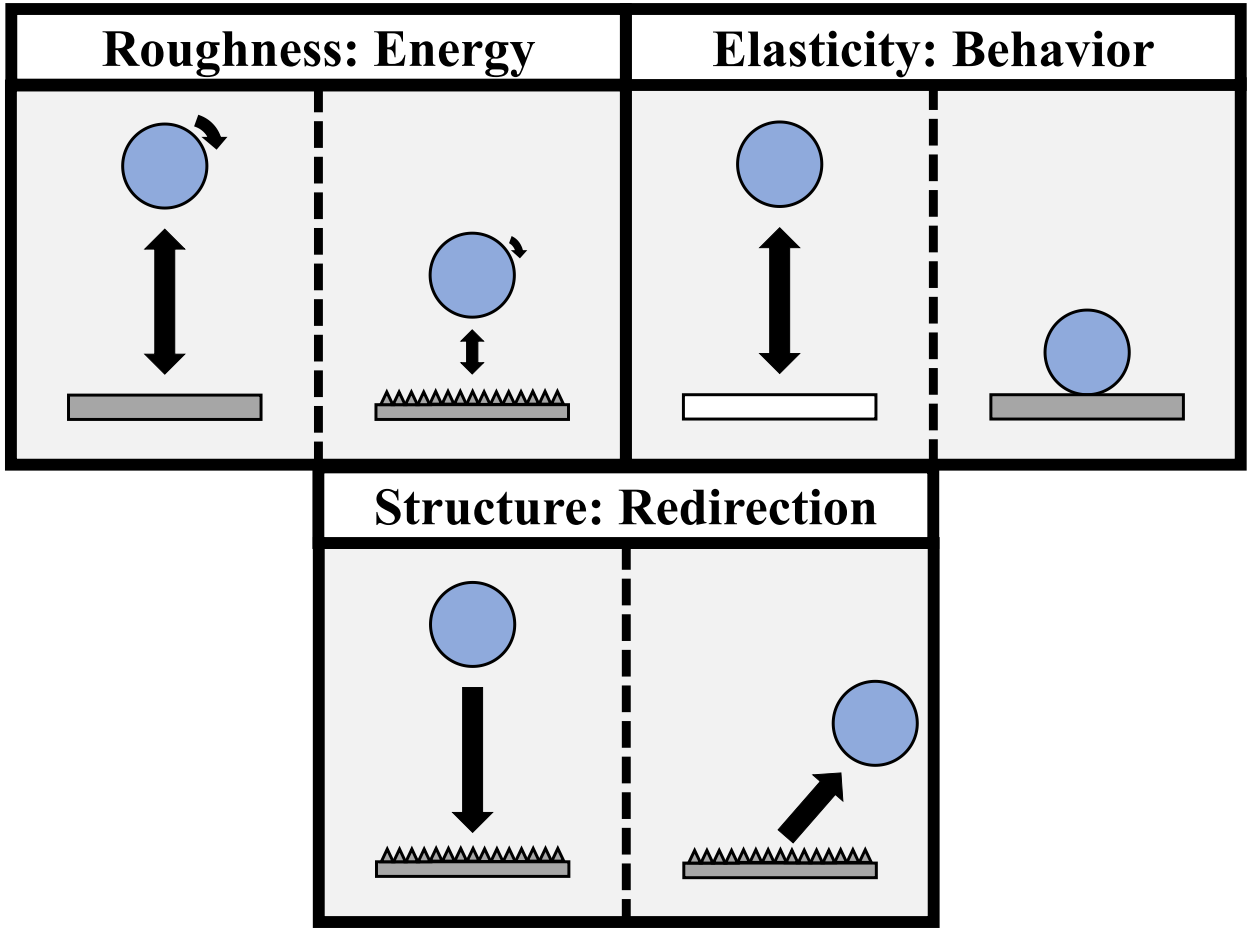

# Dynamics of Acoustically Levitated Ice Impacts on Smooth and Textured Surfaces: Effects of Surface Roughness, Elasticity, and Structure

*Adam McElligott, André Guerra, Alexandre Brailovski, Shashini Rathnayaka, Xiaodan Zhu, Alexia Denoncourt, Alejandro D. Rey, Anne-Marie Kietzig, Phillip Servio\**

Department of Chemical Engineering, McGill University, Montreal, Quebec H3A 0C5, Canada

Contents:

*S1. Derivations for the Analytical Framework*

As mentioned in the Analytical Framework section of the main text, the framework is primarily derived from two fundamental conservation equations, momentum and energy, in two dimensions with mass assumed to be conserved. In addition, these are divided into rotational and translational components. Note that these two components can only be added for the energy calculations as they are scalar terms. Therefore, there are three conservation equations in total, which are presented below.

**Momentum (Translational and Rotational):**

$$mV_1 = mV_2 + M_1 \tag{S1}$$

$$mr\omega_1 = mr\omega_2 + M_2 \tag{S2}$$

**Energy:**

$$\frac{1}{2}mV_1^2 + \frac{1}{2}m(r\omega_1)^2 = \frac{1}{2}mV_2^2 + \frac{1}{2}m(r\omega_2)^2 + E \tag{S3}$$

Here, m is the droplet mass (in kg), $V_1$ and $V_2$ are the impacting and rebounding translational velocities, respectively, with angles taken into account (in m s$^{-1}$), r is the droplet radius (in m), $\omega_1$ and $\omega_2$ are the impacting and rebounding angular velocities (in s$^{-1}$), respectively, and $M_1$, $M_2$, and E account for momentum (in kg m s$^{-1}$) and energy (in J) losses from impact. However, while present for accuracy, the latter are unnecessary for calculations.

It is now possible to define the coefficients for the system. The restitution coefficient is generally defined as the ratio between the final and initial momentum. This remains the same for this study, though it is divided into translational and rotational components, resulting in two separate coefficients.

**Translational Restitution Coefficient:**

$$\frac{M_{Translational,f}}{M_{Translational,i}} = \frac{mV_2}{mV_1} = \frac{V_2}{V_1} = \varepsilon_T \tag{S4}$$

**Rotational Restitution Coefficient:**

$$\frac{M_{Rotational,f}}{M_{Rotational,i}} = \frac{mr\omega_2}{mr\omega_1} = \frac{\omega_2}{\omega_1} = \varepsilon_R \tag{S5}$$

Next, we define two new coefficients: an energy retention coefficient, which is the ratio between the final and initial energy of the droplet, and an energy transfer coefficient, which is the ratio between the rotational energy gain and the translational energy loss. The absolute value of the latter is used for consistency as it is otherwise always negative.

**Retained Energy Coefficient:**

$$\frac{E_{Total,f}}{E_{Total,i}} = \frac{\frac{1}{2}mV_2^2 + \frac{1}{2}m(r\omega_2)^2}{\frac{1}{2}mV_1^2 + \frac{1}{2}m(r\omega_1)^2} = \frac{V_2^2 + (r\omega_2)^2}{V_1^2 + (r\omega_1)^2} = C_{Ret} \tag{S6}$$

**Transferred Energy Coefficient:**

$$\left|\frac{\Delta E_{Rotational}}{\Delta E_{Translational}}\right| = \left|\frac{\frac{1}{2}m(r\omega_2)^2 - \frac{1}{2}m(r\omega_1)^2}{\frac{1}{2}mV_2^2 - \frac{1}{2}mV_1^2}\right| = \left|\frac{(r\omega_2)^2 - (r\omega_1)^2}{V_2^2 - V_1^2}\right| = C_{Tr} \tag{S7}$$

Finally, it was necessary to differentiate between friction and deformation energy losses. This can be done by determining the average deformation energy and dividing the average energy loss on impact. The average deformation energy can be calculated using the Young-Dupré equation and the change in contact angle before and after the 40 impacts. The difference in the solid-liquid work of adhesion ($\Delta W$ in J m$^{-2}$) calculated by the Young-Dupré equation is the energy that goes to elastic potential through surface deformation. As a per-area term, it must be multiplied by the impact area (A, in m). Moreover, as the energy loss is negative, the absolute value is taken for the deformation coefficient.

**Young-Dupré Equation:**

$$W = \sigma_l(1 + cos\theta) \tag{S8}$$

**Change in Work of Adhesion:**

$$|\Delta W| = |\sigma_l(1 + cos\theta_2) - \sigma_l(1 + cos\theta_1)|$$

$$|\Delta W| = |\sigma_l[1 + cos\theta_2 - 1 - cos\theta_1]|$$

$$|\Delta W| = |\sigma_l[cos\theta_2 - cos\theta_1]| \tag{S9}$$

$$E_{Deformation} = A|\Delta W| = A|\sigma_l[cos\theta_2 - cos\theta_1]| \tag{S10}$$

**Deformation Energy Coefficient**

$$\left|\frac{E_{Deformation}}{E_{Total,f} - E_{Total,i}}\right| = \left|\frac{A\sigma_l(\cos\theta_2 - \cos\theta_1)}{\frac{1}{2}m[(V_2^2 + (r\omega_2)^2) - (V_1^2 + (r\omega_1)^2)]}\right| = C_{Def} \tag{S11}$$

Here, $\theta_1$ and $\theta_2$ are the initial and final contact angles, respectively, and $\sigma_l$ is the liquid surface tension (in J m$^{-2}$). The thermal energy (friction) loss is equivalent to $1 - C_{Def}$.

*S2. Additional SEM Images*

As mentioned in the Effects of Ice Impacts on Surface Characteristics section of the main text, the figure (**Figure S1**) below contains the SEM images for the SS and ER surfaces before and after 40 droplet impacts. As mentioned in the main text, there was no noticeable change in surface topology.

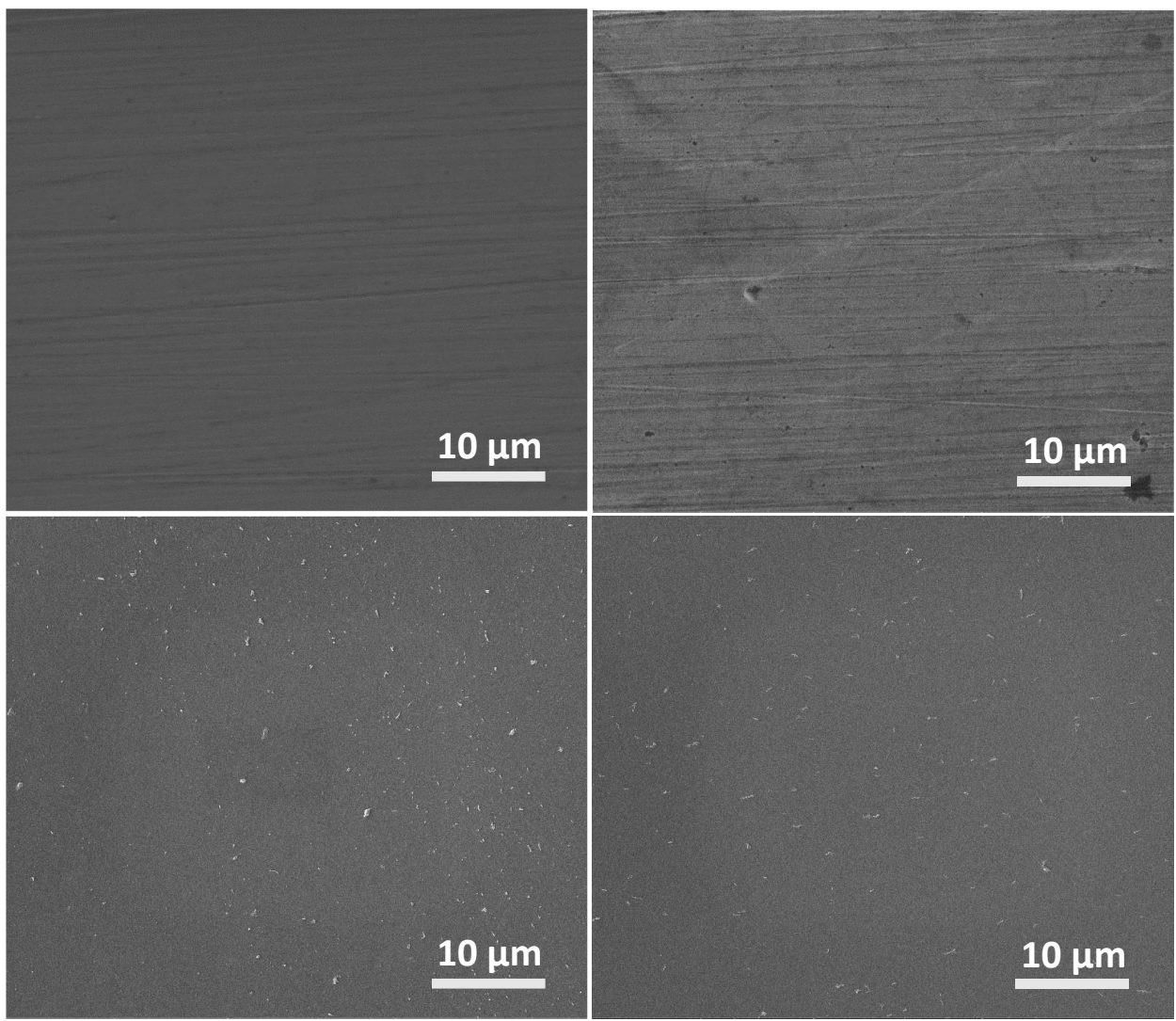

**Figure S1.** Pre- (left) and post-impact (right) surface topology for the SS (top) and ER (bottom) surfaces.

*S3. Double Rebound Case*

As mentioned in the Energy Dynamics section of the main text, the interplay between kinetic and rotational energy is displayed in runs where two rebounding events were captured. This occurred twice in the 106 runs, both on the ER surface. This makes sense as the ER surface returns more energy to the droplet than the other two stainless steel surfaces due to its elasticity. **Figure S2** shows a droplet with a greater post-impact rotational velocity, which rises only to a short height. In other words, some translational energy (in this case, 24.3 % of the lost translational energy) had been transferred to rotational energy. However, the droplet's rotational velocity is reduced after the second impact, though it reaches a much greater height. Some rotational energy (16.6% of the lost rotational energy) had returned to translational energy. This is further evidence that not all translational energy lost by the falling droplet goes to the surface. Quantifying the transferred energy is necessary to understand the system dynamics better. Moreover, it demonstrates the importance of considering the interplay between surface properties and potential secondary accretion effects. Different surfaces may vary in double rebound properties, which could be different from primary impact effects. As mentioned in the main text, a study on a larger impact area would be necessary for such an investigation.

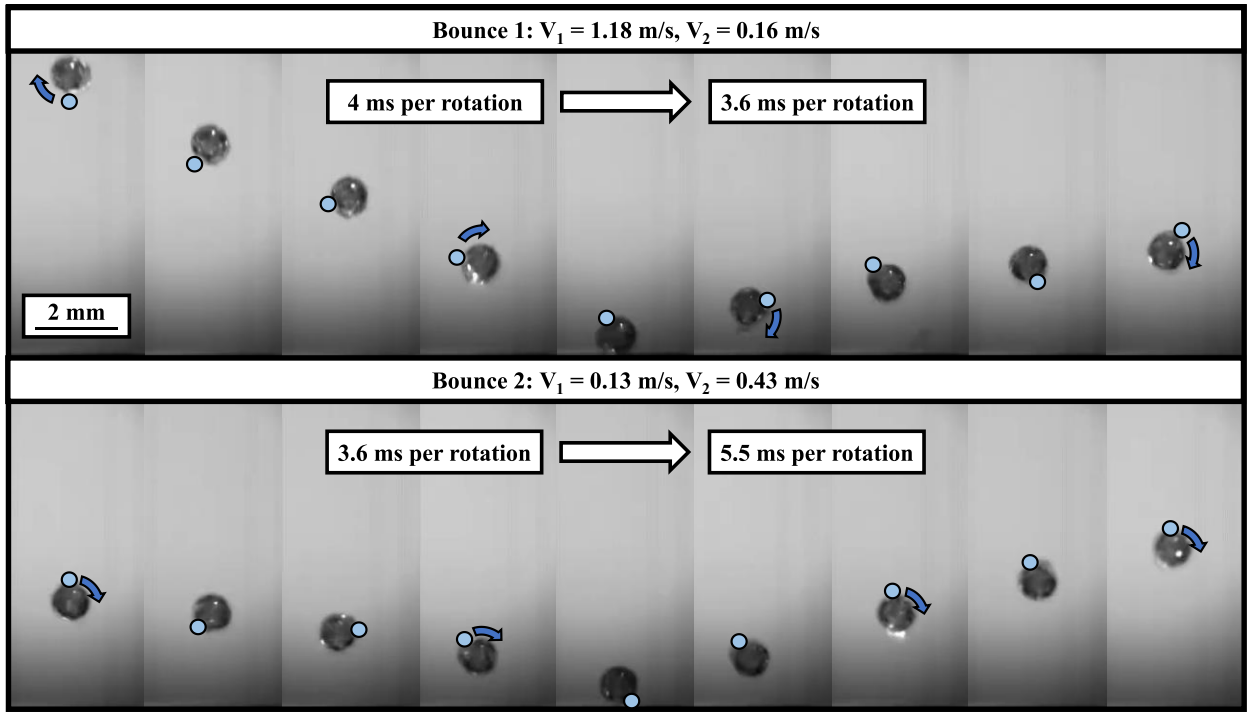

**Figure S2.** An example of a double rebound showing an exchange of translational energy to rotational, then rotational back to translational. The blue dot follows a particular point on the droplet, the arrows indicate the rotational direction, and each image is 20 frames, making small changes in rotational velocity highly noticeable via more erratic dot location.

*S4. Effect of Droplet Shape*

As mentioned in the Ice Impact Dynamics section of the main text, tt is critical to discuss how the droplet shape affects impact dynamics. As mentioned, most droplets (62%) were spherical, specifically oblate spheroids, like many atmospheric ice droplets. However, turbulence in the acoustic field, liquid-solid interfacial pressure effects, and gravity could cause a liquid bead to form at the base of the droplet, as in **Figure S3**, which would freeze and result in a rounded pyramidal structure attached to the droplet. The instances of both shapes were evenly distributed for all surfaces as they are a consequence of the levitator function only. Regarding shape, impact dynamics are affected by the impact area on the droplet. A more spherical shape will likely have a greater impact area, like the one highlighted in **Figure S3**. In contrast, adding a pyramidal structure could result in either two smaller impact points (1 and 3 in the figure) or even a single small point of impact (3) before rebounding.

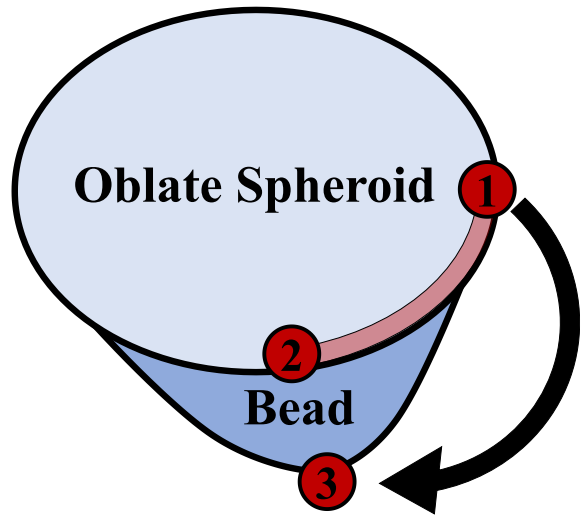

**Figure S3.** The oblate spheroid droplet shape is light blue, with the additional pyramidal structure in dark blue. Oblate spheroid friction could occur from 1 to 2 along the edge, while pyramidal droplets skip from 1 to 3 without further contact or impact on 3 only.

The larger spherical impact area is subject to more friction as the droplet interacts more with the surface, which means those droplets may exhibit reduced momentum and energy retention. This is important to consider as the spherical droplets are closer in shape to rain. If more of these droplets are present (as would be the case in the atmosphere), there could be significant wear on the surface, unlike what was found in this study. Moreover, oblate spheroid droplets spun less, resulting in less energy to overcome friction. To confirm, we can briefly examine the coefficients discussed in the main text in **Table S1**.

**Table S1.** Energy retention, energy transfer, and translational restitution coefficients for each surface separated by oblate spheroid (first) and with the pyramidal structure (second).

|     | $C_{Ret}$   | $C_{Tr}$    | $\varepsilon_T$ |
|-----|-------------|-------------|-----------------|
| SS  | 0.16 / 0.26 | 0.15 / 0.26 | 0.17 / 0.19     |
| ER  | 0.23 / 0.38 | 0.19 / 0.28 | 0.29 / 0.34     |
| LT  | 0.08 / 0.10 | 0.08 / 0.03 | 0.12 / 0.18     |

The data in the table shows no statistical difference between shapes for the LT surface. As a high-friction surface, even small contact areas can result in sufficient friction where energy retention, energy transfer, and translational restitution are similar. In addition, the values and trends for momentum retention are statistically identical. However, there is a clear difference between the energy coefficients for the other surfaces. In both cases, the pyramidal addition resulted in greater energy retention and transfer to rotational energy, noting that the trend remains identical to what was observed in the main text. This suggests that the oblate spheroid droplets were subjected to a greater frictional force, and the presence of the pyramidal addition on other droplets increased the average values of the coefficients. Moreover, impacts on areas shaped like those at point 3 in **Figure S3** are likely to enhance the rotation of the droplet and result in greater transfer from translational to rotational energy while retaining the overall energy loss behavior found generally. In summary, the droplet shape affected energy retention and energy transfer: less spherical droplets, though not as common, were subject to reduced friction and had impact areas that allowed for more post-impact rotational energy. As such, the energy values presented in this study may be slightly higher than what could be experienced in atmospheric conditions.